\documentclass[english,doublespacing,preprint,preprintnumbers,amsmath,amssymb,fleqn,epsfig,psfig]{revtex4-1}
\usepackage[T1]{fontenc}
\usepackage[latin1]{inputenc}
\usepackage{graphics}
\usepackage{graphicx}
\usepackage{multirow}
\usepackage{float}
\usepackage{bm}
\usepackage{color}   
\usepackage{dcolumn} 
\usepackage{amsmath}
\usepackage{amssymb}
\usepackage{xcolor}
\usepackage{babel}
\usepackage[twoside]{rotating}

\begin{document}

\title{A Density Functional study of Covalency in the Trihalides of Lutetium and Lawrencium}
\author{Ossama Kullie} 
\thanks{Electronic mail:kullie@uni-kassel.de}
\affiliation{Theoretical Physics, Institute for Physics, Department
of Mathematics and Natural Science, University of Kassel, Germany. Institute de Chimie de Strasbourg, CNRS et Universit\'e de Strasbourg, 
Laboratoire de Chimie Quantique, 4 rue Blaise Pascal, F--67070, Strasbourg, France}

\footnotesize

\begin{abstract}
In this work we present a four component relativistic theoretical investigation of the  trihalides  of lutetium  and lawrencium, LuX3, LrX3 (X= F, Cl, Br, I) respectively using density functional theory (DFT) with different density functional and a geometrical optimisation procedure as implemented in DIRAC-package. The results show the trend of bonding from lighter to the heavier halide atoms and between 4f/5f  atoms Lu and Lr.
\end{abstract}
\maketitle



\section{Introduction}
\begin{figure}
\begin{picture}(+400,+400)
\includegraphics[height=17cm]{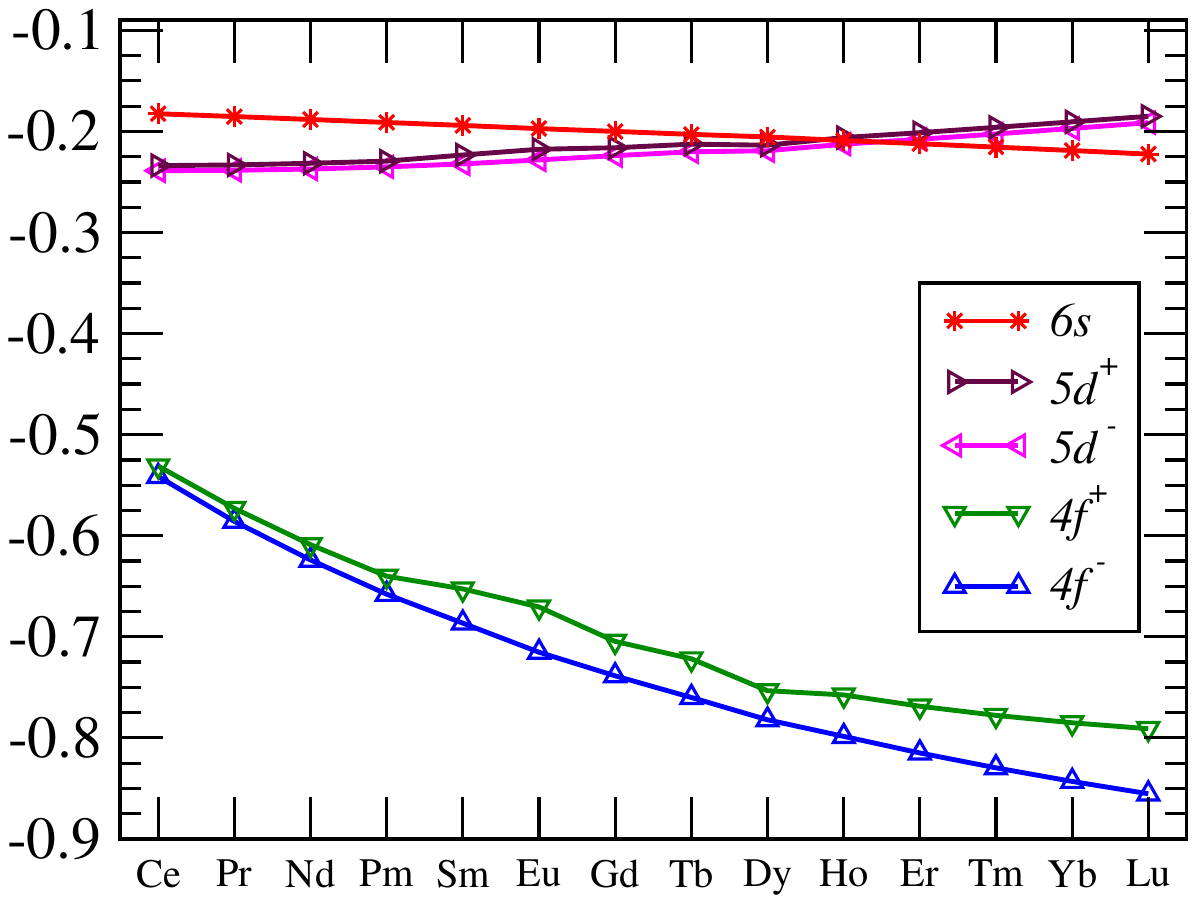}
\end{picture}
\caption{\label{fig:LnE} \footnotesize Energies (in a.u.) of the $4f$, $5d$ and $6s$ orbitals for Ce-Lu from 4-component relativistic Hartree-Fock calculations averaging over the $4f^x5d^16s^2$ (x=1, 14) valence configuration.}
\end{figure}

\begin{figure}
\begin{picture}(+300,+400)
\includegraphics[height=17cm]{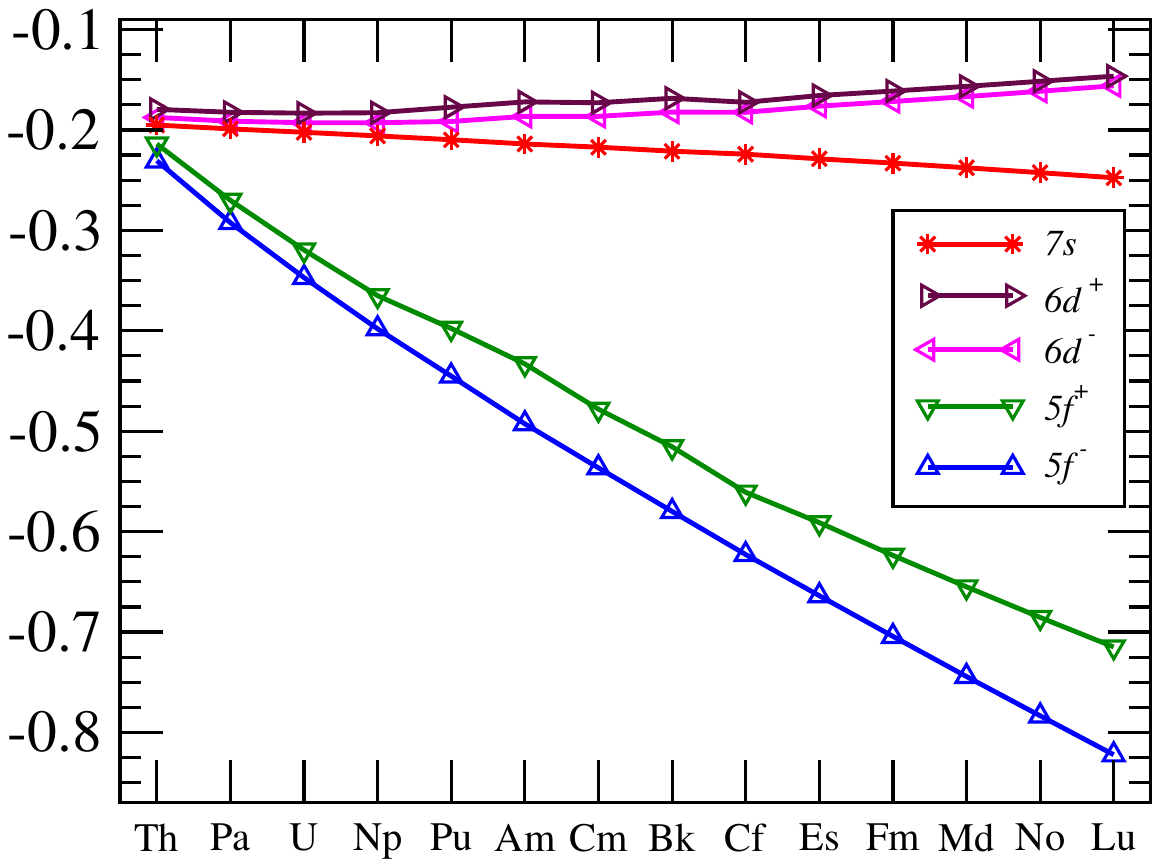}
\end{picture}
\caption{\label{fig:AnE} \footnotesize Energies (in a.u.) of the $5f$, $6d$ and $7s$ orbitals for Th-Lr
from 4-component relativistic Hartree-Fock calculations averaging over the 
$5f^x6d^17s^2$  (x=1, 14) valence configuration.}
\end{figure}

\begin{figure}
\begin{picture}(+300,+400)
\includegraphics[height=17cm]{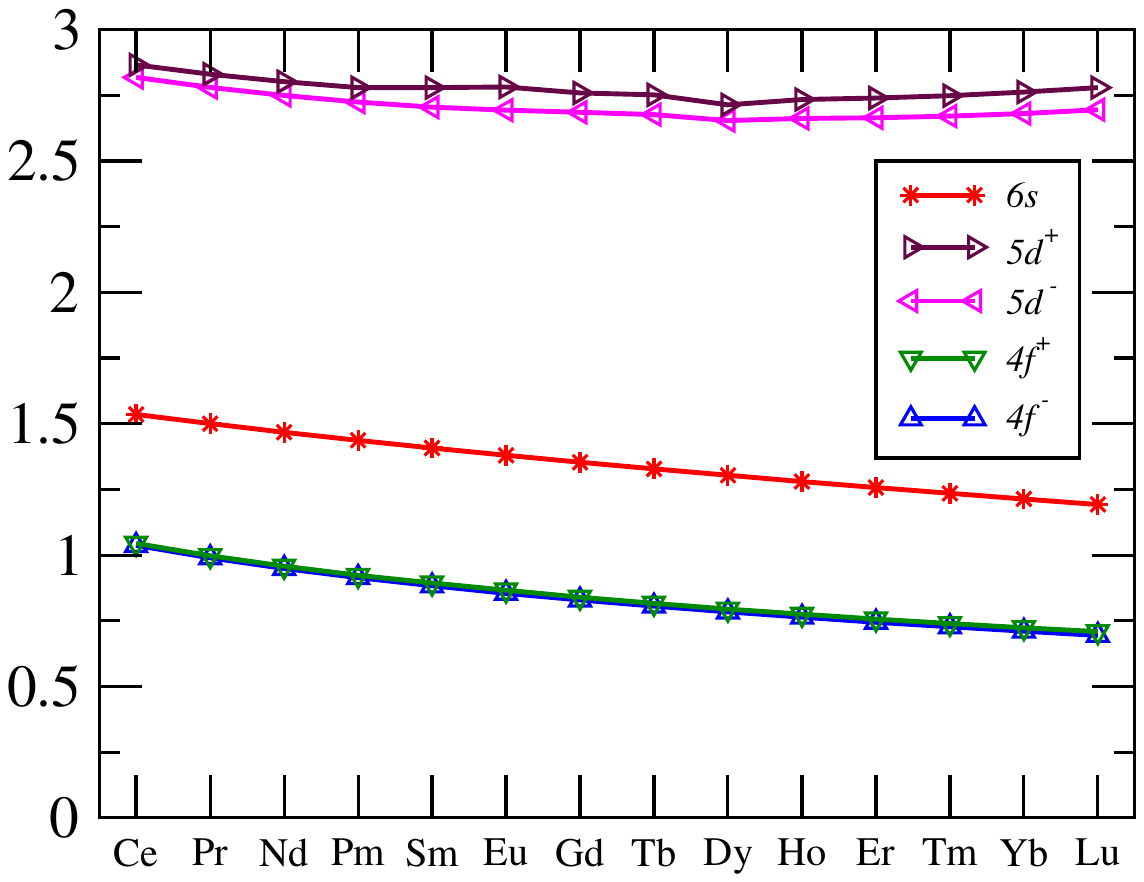}
\end{picture}
\caption{\label{fig:LnR} \footnotesize Mean radius $\left<r\right>$ (in Bohr) of the $4f$, $5d$ and $6s$ orbitals for Ce-Lu from 4-component relativistic Hartree-Fock calculations averaging over the $4f^x5d^16s^2$ (x=1, 14)valence configuration.}
\end{figure}

\begin{figure}
\begin{picture}(+300,+400)
\includegraphics[height=17cm]{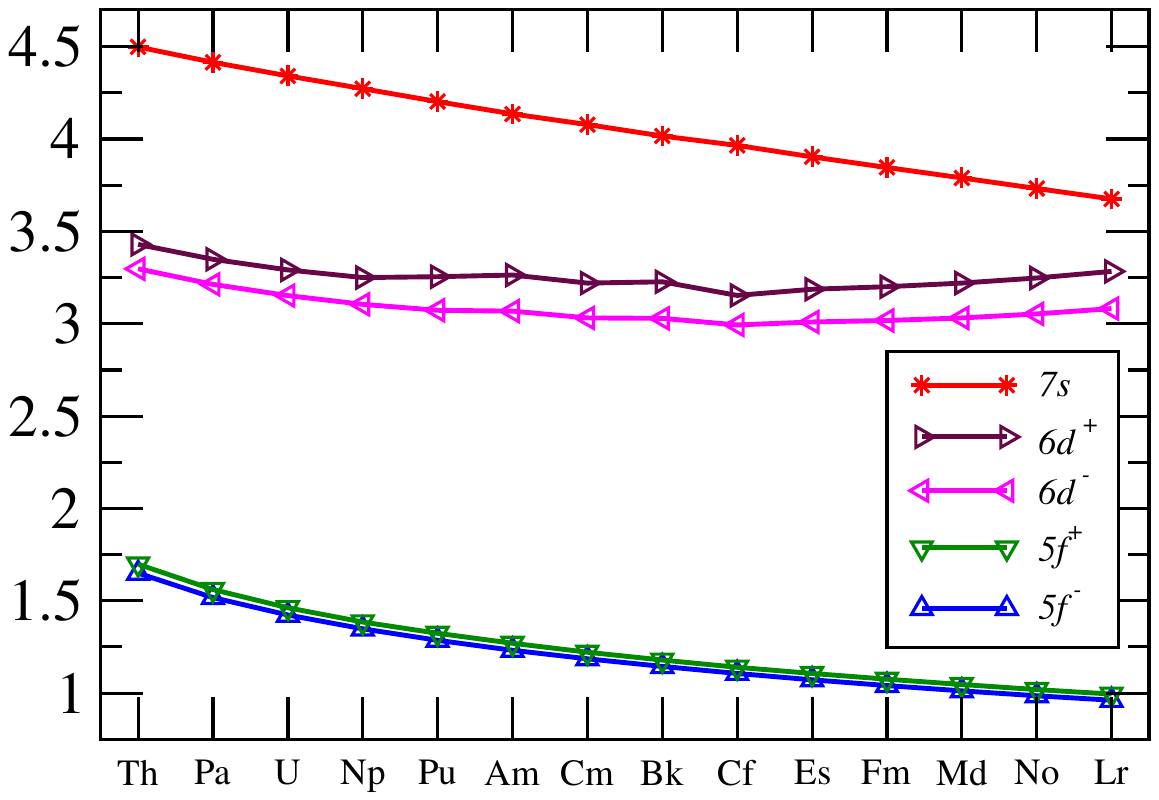} 
\end{picture} 
\caption{\label{fig:AnR} \footnotesize Mean radius $\left<r\right>$ (in Bohr) of the $5f$, $6d$ and $7s$ orbitals for Th-Lr from 4-component relativistic Hartree-Fock calculations averaging over the $5f^x6d^17s^2$ , (x=1, 14) valence configuration.}
\end{figure}

In a 1988 review in the \textit{Handbook on the Physics and Chemistry of Rare Earths} C. K. J{\o}rgensen asked the rhetorical question 'Is Quantum Chemistry feasible ?' and with special regard to these elements he answered 'Sorry, not today; perhaps next century' \cite{jorgensen:lanthanides}. Presently, in the new century,  the structure and reactivity of $f$ elements is a flourishing domain of theoretical chemistry [INSERT reviews..]
The main features of the chemistry of the lanthanides and the actinides can be deduced from simple atomic calculations.
In Figs. \ref{fig:LnE}-\ref{fig:AnR} we present orbital properties extracted from numerical 4-component relativistic
Hartree-Fock calculations\cite{grasp} averaging over the valence configuration $(n-2)f^x(n-1)d^1ns^2, (x=1,14)$ 
for the neutral lanthanide and actinide atoms. We have chosen this configuration, which is not the ground state 
configuration for all the $f$ elements, since it gives access to information about the $(n-2)f$, $(n-1)d$ and $ns$ orbitals.
From the orbital energies in Fig. \ref{fig:LnE} we observe a distinct energetical separation of the $4f$ orbitals
from the $5d$ and $6s$ orbitals of the lanthanides which explains their chemistry dominated by the $+3$ oxidation state.
We note that the $5d$ levels cross the $6s$ level towards the end of the series, a feature which may induce convergence
problems in atomic calculations not exploiting the full atomic symmetry. Fig. \ref{fig:AnE} shows a somewhat different
situation for the actinides in that the $5f$ levels are energetically close to the $6d$ and $7s$ levels at the
beginning of the series, but are then strongly stabilized towards the end of the series. These features translate into
a rich oxidation chemistry for the early actinides and a restriction to the $+3$ oxidation state for the late actinides.
This in turn explains the challenge of separating the minor actinides americium and curium from the lanthanides in
the treatment of nuclear waste [REF].
Looking at mean radii $\left<r\right>$, we observe  in Fig. \ref{fig:LnR} a distinct spatial separation of the $4f$, 
$5d$ and $6s$ orbitals in the lanthanides, whereas the $(n-1)d$ and $ns$ orbitals come quite a bit closer in the
actinides. [DISCUSS lanthanide contraction] 
It is also interesting to observe that whereas the spin-orbit splitting in the $(n-2)f$ shell is
considerably stronger than for the $(n-1)d$ shell, any difference in spatial extent is hardly visible for the 
spin-orbit components of the $(n-2)f$ shell. This can be understood from the fact that the $(n-2)f$ orbitals are
in general quite contracted and so any deformation of the orbitals is energetically very much more expensive than
for the $(n-1)d$ shell.

With these observations in mind it was all the more surprising when Clavagu{\'e}ra \textit{et al.} \cite{dognon:2005} reported a 
clear example of $4f$ participation in bonding in LuF$_3$ since lutetium is at the very end of lanthanide series
where one would expect the $4f$ orbitals to be the most inert. The conclusion was seriously questioned by Roos \textit{et al.}
\cite{roos:2008} as well as Ramakrishnan \textit{et al.} \cite{ramak:2009}. 

The purpose of the present paper is to investigate the possible $4f$ participation in LuF$_3$ by an independent approach. 
We employ a trick that may be useful for other purposes as well.
We extend our electronic structure analysis to all trihalides (X=F, Cl, Br, I) of lutetium as well as of lawrencium.
We thereby provide a comparison of covalency between these two elements, of relevance for the delicate problem of separation
of the late actinides from the lanthanides. The paper is organized as follows: In section \ref{sec:theory} we describe our
methodological approach. Computational details are given in section \ref{sec:comp}. In section \ref{sec:results} we present
and discuss the results of our geometry optimizations and electronic structure analysis, before concluding in section \ref{sec:conclusion}.

\section{Theoretical considerations}\label{sec:theory}
The question about the participation of $(n-2)f$ orbitals in bonding in the lanthanides and actinides is very much a
leading question:
\begin{enumerate}
\item It assumes that one can identify these atomic orbitals in the electronic structure of the
molecule.
\item It assumes that one can unambiguously distinguish bonding from non-bonding contributions to 
the electronic structure of the molecule. 
\end{enumerate}
In order to tackle the first point we perform 4-component relativistic Hartree-Fock (HF)
and Kohn-Sham (KS) calculations of the trihalides of lutetium and lawrencium (LuX$_3$ and LrX$_3$, X=F, Cl, Br and I) and investigate their electronic
structure by projection analysis\cite{saue:H2Ohom}, that is we expand the molecular orbitals (MOs) in pre-calculated 
orbitals (index $j$) of the constituent atoms (index $A$). 
\begin{equation}
\psi_i^{MO}=\sum_{Aj}\psi_j^{A}c^{A}_{ji} + \psi_i^{pol}.
\end{equation}
The fragment orbitals are usually restricted to the occupied orbitals of the selected configurations of the 
constituent atoms of the molecule. The expansion is completed by the polarization contribution $\psi_i^{pol}$
which by construction is orthogonal to the fragment orbitals. Once the expansion coefficients $c^{A}_{ji}$ 
have been obtained a population analysis may be carried out completely analogous to, but without the basis set
sensitivity of a Mulliken population analysis. The selection of fragment orbitals should be adjusted if the
gross population of the polarization contribution is significant.

The next point is somewhat more difficult due to the invariance of the electronic 
energy under rotations of the occupied orbitals of these closed-shell molecules. This rotational freedom can
be exploited to transform from canonical Hartree-Fock or Kohn-Sham orbitals to localized orbitals, or, in the 
terminology of Mulliken\cite{mulliken:nobel}, from spectroscopic to chemical MOs. However, there is an abundance
of possible localization criteria and thus perhaps not a clear answer from such an approach. The problem would
have been simpler for the trihalide of any other lanthanide than lutetium since one would then expect inert $4f$ orbitals to
form an open shell. There would accordingly be a \textit{variance} of the electronic energy with respect to 
rotations between the open- and closed-shell orbitals. However, this immediately suggests the strategy that 
we will pursue in this paper: We will place $4f$ orbitals from an atomic calculation in an artificial open-shell 
with fractional occupation close to, but not equal to 14. If the $(n-2)f$ atomic orbitals now localize to 
the open shell, we have unambiguous evidence that they are chemically inert.

\section{Computational details}\label{sec:comp}
All calculations reported in this paper are based on the 4-component Dirac-Coulomb Hamiltonian with a Gaussian charge 
distribution as the nuclear model using the recommended values of Ref.~\cite{Visscher_Dyall_1997}.
The numerical atomic calculations were performed using the GRASP code \cite{grasp}. 
Molecular relativistic 4-component Hartree-Fock and density functional (PBE \cite{Perdew_Burke_Ernzerhof_1996, Perdew_Ernzerhof_1996}  
and B3LYP \cite{Stephens_Devlin_Chabalowski_Frisch_1994,Becke_1993,Lee_Yang_Paar_1988}) calculations
were carried out using the DIRAC08 package \cite{DIRAC08}.
We employed the cc-pVTZ Gaussian basis sets of Dunning and co-workers \cite{dunning:1989,dunning:1993,dunning:1999} 
for F, Cl and Br and equivalent sets of Dyall and co-workers \cite{basis_Dyall_4p-6p,basis_Dyall_4d_5d_5f,basis_dyall_4f} 
for I, Lu and Lr. The small component basis set for the 4-component relativistic calculations has been generated 
using restricted kinetic balance imposed in the canonical orthonormalization step\cite{Visscher_Saue_2000}.
All basis sets are used in uncontracted form. HF and KS geometry optimizations were carried out using analytic
and numerical gradients, respectively. For pyramidal ($C_{3v}$) and planar ($D_{3h}$) structures we employed
the lower C$_s$ and C$_{2v}$ symmetries, respectively. Test calculations with cc-pVDZ basis sets indicate that the
resported structures can be considered converged with respect to the chosen basis sets. For the projection analysis
fragment orbitals were generated by average-of-configuration HF calculations and KS calculations with fractional occupation, corresponding to ground state electronic configurations of the atoms.

\section{Results and discussion}\label{sec:results}
\subsection{Molecular structure}
In table \ref{tab:LuX3}, \ref{tab:LrX3} we present  HF, B3LYP and PBE calculation for lutetium and lawrencium trihalides  LuX$_3$, LrX$_3$ (X=F, Cl, Br, I) respectively. For the former trihalides, the LuX$_3$, there exist a large number of theoretical works whereas very limited works concerned the lawrencium trihalides. Since we will discuss the two table separately. In table \ref{tab:LuX3} we summarise our result together with theoretical and  experimental values from the literature, given also in the table the method (or the Hamiltonian) and the basis set used to obtain these values. Firs of all looking to the bond lengths in table \ref{tab:LuX3} on see that our values using B3lYP have the best agreement with experimental ones and second there is a good agreement with the literature for the respective method (or Hamiltonian), especially with the HF values of ref. \cite{lanza:2005},\cite{weigand:2009} and the PBE of ref. \cite{dognon:2005}.\\
Concerning the geometrical shape and the bond angle we see that HF give a planer geometry for all the trihalides whereas B3LYP and PBE gives a pyramidal geometry for the LuF$_3$ only and a planar geometry for the other trihlides. Increasing of the bond angle towards the heavier trihlides can be understood having in mind that the heaver halides have weaker ability to polarize the metal atom which means less distribution of the electronic density around the metal atom favoring a planar geometry with the highest bond angel of 120 grad and longer bond length towards the heavier halides as seen in table \ref{tab:LuX3}.\\
\subsection{$4f$ orbital participation in bonding in LuF$_3$}
We have studied the possible $4f$ orbital participation in bonding in LuF$_3$ by the approach presented in sections \ref{sec:theory}
and \ref{sec:comp}. The results are summarized in table \ref{tab:LuF3}. We first calculated the neutral lutetium atom in $D_{2h}$ 
with linear supersymmetry. The resulting coefficients were exported to $C_1$ symmetry. The Lu $4f$ orbitals were next imported 
into a molecular calculation on LuF$_3$ in the optimized pyramidal geometry and kept frozen in an initial calculation on the 
neutral molecule. Starting from the resulting molecular coefficients a series of KS calculations with fractional occupation, 
that is 14-$\delta$ electrons ($\delta \in \left[0.0,0.2\right]$) in 14 spinors, were carried out and the resulting molecular 
orbitals studied by projection analysis. In each calculation vector for each occupied orbital was selected based on overlap
with the starting molecular coefficients. However, it is seen from table \ref{tab:LuF3} that the electronic energy goes
smoothly into the energy of the fully relaxed neutral molecule as the hole $\delta$ tends towards zero, indicated that we
indeed obtained the ground state of the molecule for each value of $\delta$.\\\\
{\bf Conclusions}\label{sec:conclusion}
The presented four component relativistic result for the trihalides of lutetium and lawrencium, LuX3, LrX3 (X= F, Cl, Br, I) 
respectively using density functional theory (DFT) shows that 
the trend of bonding is from lighter to the heavier halide atoms and between 4f/5f  atoms Lu and Lr.


\newpage
\begin{figure}
\begin{picture}(+300,+400) 
\includegraphics[height=17cm]{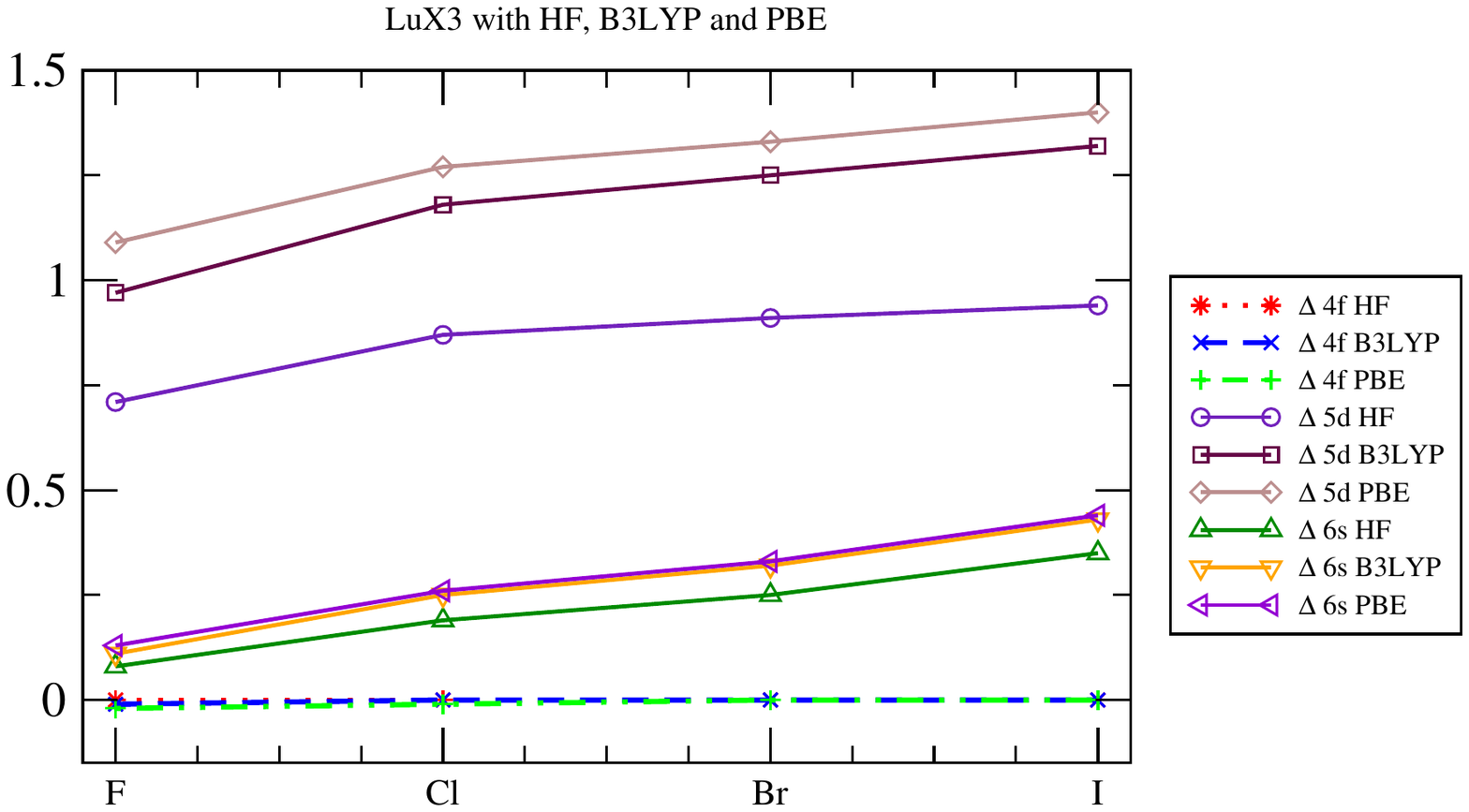}
\end{picture} 
\caption{\label{fig:LuX3-HFB3LYPPBE} \footnotesize LuX${_3}$ using HF,B3LYP- and PBE- functional, change of population with respect to 4f$^{14}$ 5d$^{0}$ 6s$^{0}$ configuration}
\end{figure}
\begin{figure}
\begin{picture}(+300,+400) 
\includegraphics[height=17cm]{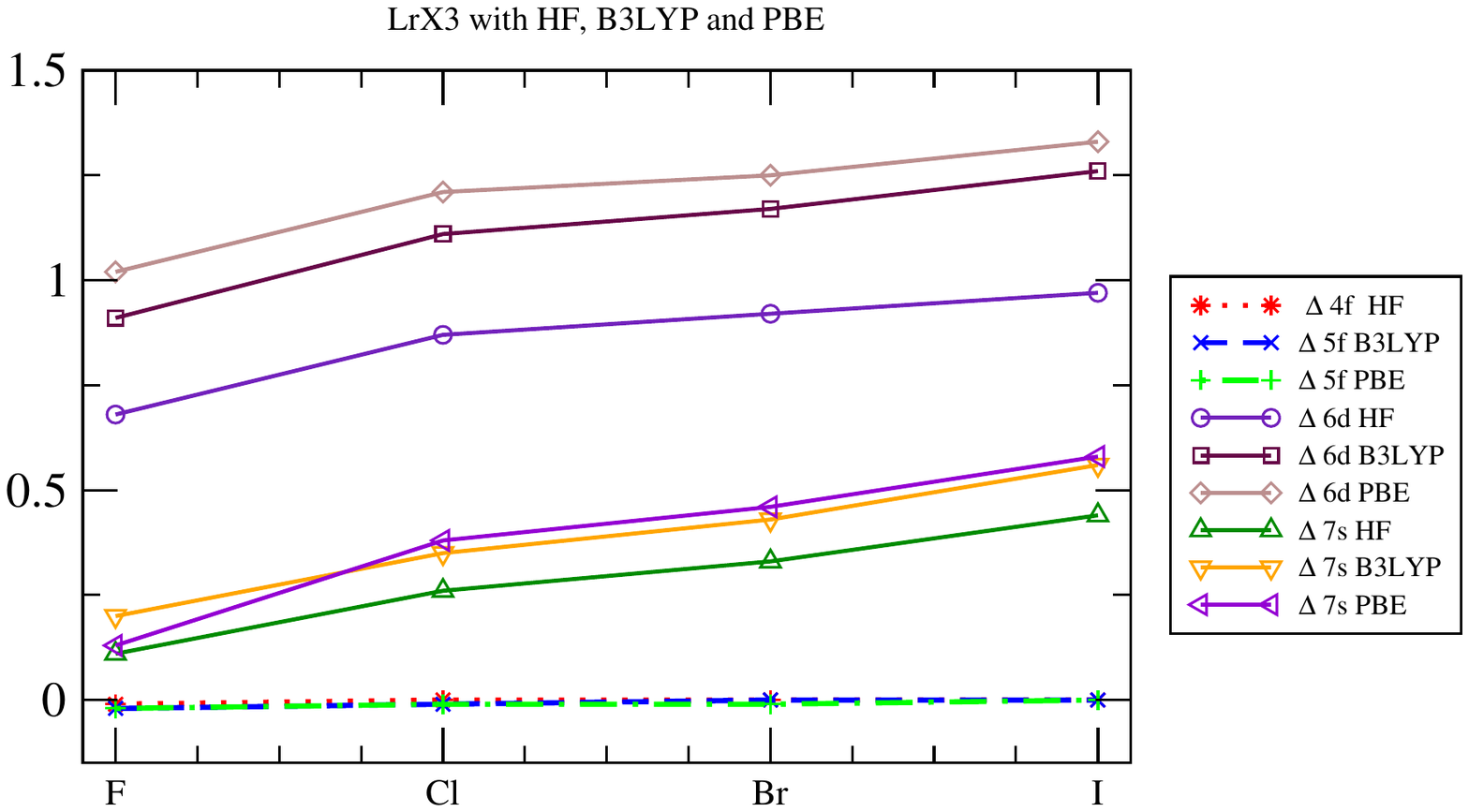}
\end{picture} 
\caption{\label{fig:LrX3-HFB3LYPPBE} \footnotesize LrX${_3}$ using HF ,B3LYP- and PBE- functional, change of population with respect to 5f$^{14}$ 6d$^{0}$ 7s$^{0}$ configuration}
\end{figure}
\clearpage
{\footnotesize
\begin{sidewaystable}
\caption{\label{tab:LuX3} Geometric parameters of lutetium trihalides LuX$_3$}
\begin{ruledtabular}
\begin{tabular}{llll|llll|llll}
          &           &                         &                    &\multicolumn{4}{c|}{Lu-X bond length (in \AA)}  & \multicolumn{4}{c}{X-Lu-X bond angle (in $^\circ$)}  \\
Method    &Hamiltonian&Basis                    &  Ref               & LuF$_3$ & LuCl$_3$ & LuBr$_3$ & LuI$_3$  & LuF$_3$ & LuCl$_3$ & LuBr$_3$ & LuI$_3$\\\hline
 HF       &DC         &cc-pVTZ                  & \bf pw             & \bf 1.980 & \bf 2.440 & \bf 2.592& \bf 2.821 & \bf 120  &\bf  120 & \bf 120  & \bf 120\\
 HF       &RECP       & (TZ2D)                  & \cite{lanza:2005}  & 1.984 & 2.439 &  --    &  -- &   --  &   --    &   --     &   \\
 HF       &RECP/LPP   &                         & \cite{weigand:2009}& 1.984 &    --    &    --     &   --   &  120  & --  &  --    &  --   \\
 B3LYP    &DC         &cc-pVTZ                  & \bf pw             &  \bf 1.972 & \bf 2.407 & \bf 2.557& \bf 2.777 & \bf 118  & \bf 120 &\bf  120  & \bf 120\\
 B3LYP    &RECP       &\cite{cundari:1993}+pol  & \cite{vetere:2000} & 1.991 & 2.447 & 2.590  & 2.809& 118.9& 120 & 120  & 120 \\
 B3LYP    &DKH        &ANO-RCC                  & \cite{roos:2008}   & 1.985 &   --  & --     & --   &  120  &  -- &  --  & -- \\
 B3LYP    &RECP       &                         & \cite{Joubert:1998}& 1.97  & 2.44  & --     & --   &  118.4&119.9&  --  & --\\
 PBE      &DC         &cc-pVTZ                  & \bf pw             &  \bf 1.965& \bf 2.393 & \bf 2.543 & \bf2.761 & \bf 115.6 &\bf  120 &\bf  120  & \bf 120\\
 PBE      &ZORA       &(QZ4P)                   & \cite{dognon:2005} & 1.969 & 2.396 & 2.544  & 2.779 & 116.2$^{P}$& 120  & 120  &120 \\
 PBE0     &RECP       &\cite{cundari:1993}+pol  & \cite{vetere:2000} & 1.995 & 2.443 & 2.586  & 2.804 & 118.0& 120 & 120  & 120\\
 PBE+U    &AE DKH     & LCGTO+FF+DF             &\cite{ramak:2009}   & 1.972 &  --   & --     & --    & 117.7& --  & --   & -- \\
 PBE0     &RECP       &\cite{cundari:1993}+pol  & \cite{vetere:2000} & 1.983 & 2.429 & 2.571  & 2.790  & 119.0& 120 & 120  & 120 \\
 PBE0     &RECP       &\cite{dolg:1989}         & \cite{adamo:2000}  & 1.977 & 2.419 & 2.566  & 2.785 & 119.3& 120 & 120  & 120\\
 CCSD(T)  &RECP       &(TZ2D)                   & \cite{lanza:2005}  & 1.974 & 2.413.& --     & --    & 120  & 120 & --   & --\\
 CCSD(T)  &RECP/ LPP  &                         &\cite{weigand:2009} & 1.982 &       &        &     &  120  & --  &  --    &  ---    \\
 CASPT2   &DKH        &ANO-RCC                  & \cite{roos:2008}   & 1.961 &  --   &  --    & --   &  120  & --  &   --&  --    \\
 MP4      &RECP       &\cite{cundari:1993}+d    & \cite{sol:2005}    & 1.976 & 2.379 & 2.536  & 2.749 & --   & --  &  --  & --\\
 MP2      &RECP       &(TZ2D)                   & \cite{lanza:2005}  & 1.966 & 2.397 &  --    &--     &  --  &     &  --  &  \\
Rec. &                &                         & \cite{kovacs:2004} & 1.943 & 2.373 & 2.516 & 2.733  & 116.0 & 120 & 120  & 120 \\\hline
Exp. (e-values)  &&  & \cite{zasorin:1988} & -- & 2.374(10) & 2.516(8) & 2.735(6) & --& 114.5(24) & 117.2(17)  & 116.6(10) \\
Exp. (g-values)  &&  & \cite{zasorin:1988,hargittai:1988} & 1.968(10) & 2.417(6)  & 2.557(4) & 2.768(3) & --& 111.5(20) & 115.0(11)& 115.6(6) \\
Exp. (g-values)    &&  & \cite{giricheva:2009,giricheva:2000} & -- & 2.403(5)& 2.553(5)& --& --& 117.9(1.3)& 115.3(10)& --\\
\end{tabular}
\end{ruledtabular}
{\tiny
pw:present work. 
$^{RECP}$:Relativistic effective core potential; P:\cite{Pyykko_P} note the value is erroneously given in \cite{dognon:2005};
Exp.(e-values) and Exp.(g-values): Equilibrium and thermally averaged values respectively;
d:augmented with diffuse function on X-atoms;
AE: All electron are correlated in MP2 calculation; 
DKH: Douglas-Kroll-Hess;
RCC:Relativistic  (semi-)core correction;
pol: Diffuse function was added by the authors to the basis set of ref.\cite{cundari:1993};
LPP+CPP: 5f-in-core large core pseudopotential (LPP), see \cite{weigand:2009}.
Rec: Recomended values by the authors of ref \cite{kovacs:2004}.}
\end{sidewaystable}}
\clearpage
\begin{table}
\caption{\label{tab:LrX3} Geometric parameters of lawrencium trihalides LrX$_3$}
\begin{ruledtabular}  
\begin{tabular}{llll|llll|llll}
           &             &         &                                & \multicolumn{4}{c|}{Lr-X bond length (in \AA)}  & \multicolumn{4}{c}{X-Lr-X bond angle (in $^\circ$)}  \\
 Method    & Hamiltonian &Basis    &  Ref                           & LrF$_3$ & LrCl$_3$ & LrBr$_3$ & LrI$_3$ & LrF$_3$ & LrCl$_3$ & LrBr$_3$ & LrI$_3$\\\hline
 HF        & DC          &cc-pVTZ  & pw                             & 2.024 & 2.474  & 2.624   & 2.850 & 117   & 120   & 120  & 120\\
 HF        & RECP/LPP    &         &\cite{moritz:2007,weigand:2009} &   2.037  &    --    &    --  &  --     & 120     &-- &-- &  \\
 B3LYP     & DC          &cc-pVTZ  & pw                             & 2.013 & 2.443  & 2.597   & 2.815 & 110.6 & 116.6 & 120  & 120\\
 PBE       & DC          &cc-pVTZ  & pw                             & 2.010 & 2.428  & 2.583 & 2.80  & 109 & 114.5 & 120  & 120\\
CCSD(T)    & RECP/LPP    &         & \cite{weigand:2009}            & 2.020 &  --  & --  &   --  &  114.9 & --  &  --    &  ---    \\
\end{tabular}
\end{ruledtabular}  
\end{table}  
LPP = 5f-in-core large core pseudopotential (LPP), see \cite{weigand:2009}.\\ 
\clearpage
\begin{table}
\caption{\label{tab:LuF3} Summary of PBE calculations on LuF$_3$ with a fictitious hole of $\delta$ electrons.
The total electronic energy is -14880$+\Delta E$ $E_h$. Total charge Q(Lu) on the lutetium atom as well as valence orbital populations, including $4f$ occupation $4f^{open}$ of the open shell, from projection analysis are also given.}
\begin{ruledtabular}
\begin{tabular}{c|cc|ccccc}
14-$\delta$ & $\Delta E$($E_h$) & Q(Lu) & 4f & 4f$^{open}$ & 5p & 5d & 6s\\\hline
13.80 & -0.750 & +1.73 & 13.80 & 13.0 & 6.00 & 1.1 & 0.1\\
13.85 & -0.773 & +1.69 & 13.86 & 12.3 & 6.00 & 1.1 & 0.1\\
13.90 & -0.795 & +1.65 & 13.91 & 10.8 & 6.00 & 1.1 & 0.1\\
13.95 & -0.815 & +1.62 & 13.95 & 9.4  & 6.00 & 1.1 & 0.1\\
13.98 & -0.826 & +1.61 & 13.97 & 8.9  & 6.00 & 1.1 & 0.1\\\hline
14.00 & -0.833 &       &       &     &     &    \\
\end{tabular}
\end{ruledtabular}
\end{table}
\begin{table}
\caption{\label{tab:pop}{\footnotesize Populations of the orbitals ns, (n-1)d and (n-2)f for Lutetium (n=6) and  Lawrencium (n=7) atoms in the trihalides molecules LuX3 and LrX3 (X=F, Cl, Br,I) using HF, B3LYP- and PBE-functionals and cc-pVDZ basis set. LuF3, LuCl3 and  LrF3, LrCl3 are calculated in C$_s$-symmetry whereas LuBr3, LuI3 and  LrBr3, LrI3 in C$_{2v}$-symmetry.}}
\begin{ruledtabular}
\begin{tabular}{ll|llll|llll}
 &  & \multicolumn{4}{c|}{LuX3}  & \multicolumn{4}{c}{LrX3}  \\
Method &  Orbital & LuF3 & LuCl3 & LuBr3 & LuI3 & LrF3 & LrCl3 & LrBr3 & LrI3\\\hline
 HF    & (n-2)f & 14.00 & 14.0  & 14.0 & 14.0 & 13.99 & 14.0  & 14.0  & 14.0 \\
       & (n-1)d & 0.71  & 0.87  & 0.91 & 0.94 & 0.68  & 0.87  & 0.92  & 0.97\\
       & ns     & 0.08  & 0.19  & 0.25 & 0.35 & 0.11  & 0.26  & 0.33  & 0.44\\\hline
 B3LYP & (n-2)f & 13.99 & 14.0  & 14.0 & 14.0 & 13.98 & 13.99 & 14.0  & 14.0\\
       & (n-1)d & 0.97  & 1.18  & 1.25 & 1.32 & 0.91  & 1.11  & 1.17  & 1.26\\
       & ns     & 0.11  & 0.25  & 0.32 & 0.43 & 0.20  & 0.35  & 0.43  & 0.56\\\hline
 PBE   & (n-2)f & 13.98 & 13.99 & 14.0 & 14.0 & 13.98 & 13.99 & 13.99 & 14.0\\
       & (n-1)d & 1.09  & 1.27  & 1.33 & 1.40 & 1.02  & 1.21  &  1.25 & 1.33\\
       & ns     & 0.13  & 0.26  & 0.33 & 0.44 & 0.23  & 0.38  & 0.46  & 0.58\\
\end{tabular}
\end{ruledtabular}
\end{table}
\clearpage
\bibliographystyle{/usr/share/texlive/texmf-dist/tex/latex/revtex-1/apsrev4-1.bst}
\newcommand{\Aa}[0]{Aa}
%
\end{document}